\documentclass[twocolumn]{aastex631}

\usepackage{comment}
\usepackage{multirow}
\usepackage{graphicx}
\usepackage{makecell}
\usepackage[inkscapelatex=false]{svg}
\usepackage{booktabs}
\usepackage{array}
\usepackage{tabularx}

\shorttitle{RAIRS spectra of CO$_2$}
\shortauthors{Suhasaria et al.}
\graphicspath{{./}{figures/}}

\begin{document}

\title{CO$_2$ infrared spectra on silicate dust grain analogs: Implications for JWST observations}

\correspondingauthor{Tushar Suhasaria}
\email{suhasaria@mpia.de}

\author[0000-0002-0786-7307]{Tushar Suhasaria}
\affiliation{Max Planck Institute f\"ur Astronomie, K\"onigstuhl 17, 69117, Heidelberg, Germany}

\author{Vanessa Leuschner}
\affiliation{Max Planck Institute f\"ur Astronomie, K\"onigstuhl 17, 69117, Heidelberg, Germany}
\affiliation{Department for Physics and Astronomy, University of Heidelberg, Im Neuenheimer Feld 226, 69120, Heidelberg, Germany}

\author{Cornelia J{\"a}ger}
\affiliation{Max Planck Institute f\"ur Astronomie, K\"onigstuhl 17, 69117, Heidelberg, Germany}
\affiliation{Laboratory Astrophysics Group at the Friedrich Schiller University Jena, \\ Institute of Solid State Physics, Helmholtzweg 3, 07743, Jena, Germany}

\author{Caroline Gieser}
\affiliation{Max Planck Institute f\"ur Astronomie, K\"onigstuhl 17, 69117, Heidelberg, Germany}

\author{Thomas Henning}
\affiliation{Max Planck Institute f\"ur Astronomie, K\"onigstuhl 17, 69117, Heidelberg, Germany}

\begin{abstract}
Carbon dioxide is one of the three most abundant species within the ice mantles around dust grains inside molecular clouds. Since a substantial amount of interstellar grains is made of siliceous materials, we have studied the infrared profile of CO$_2$ deposited on top of a bare and ice-coated amorphous silicate (MgFeSiO$_4$) film using reflection absorption infrared spectroscopy (RAIRS). In contrast to a metal surface, the CO$_2$ IR profile shows a relaxation of the metal surface selection rule in the presence of the bare MgFeSiO$_4$ dust grain analog, which brings the IR profile closer to the observational spectra while maintaining the sensitivity of RAIRS. Experiments with the underlying CO and CH$_4$ ices show that their presence facilitates structural changes toward crystalline ice for the deposited CO$_2$ at much lower temperatures than on the polar ice layers. Warming-up experiments of CO$_2$ showed that it tends to stay on the silicate surface for much longer than on the gold surface without the silicate layer. We noticed for the first time a split in the $^{13}$CO$_2$ IR feature on the pure or ice-covered silicate grain as a marker for the onset of diffusion. The laboratory $^{13}$CO$_2$ profile then closely resembles recent JWST observations of this feature around young and embedded protostars, suggesting that it can be linked to the observed feature.
\end{abstract}

\section{Introduction} 
In cold, dense regions of molecular clouds, before star formation begins, molecules like H$_2$O, NH$_3$, and CH$_4$ form on submicron-sized interstellar dust grains through grain surface chemistry. These molecules accumulate as ice layers, creating the first polar phase, which also contains CO$_2$. CO, formed in the gas phase, subsequently accretes on top of the polar phase, forming the apolar phase \citep{boogert2015}. Additional molecules such as CH$_3$OH and CO$_2$ also form in this apolar phase, which means that CO$_2$ can be present in both phases \citep{pontoppidan2008}. H$_2$O is the main ice component and, although it is mostly amorphous at low temperatures, it can undergo a structural transition to a cubic crystalline form as molecular clouds evolve \citep{jenniskens1994}.

CO$_2$ ice, with an abundance of 10-20\% relative to H$_2$O in dense regions of low-mass star-forming clouds, as revealed in the mid-infrared by recent JWST observations \citep{mcclure2023}, is a major carbon-bearing species. Together with CO, it contributes significantly to the overall carbon budget, providing a vital source of elemental carbon for the formation of complex organic molecules. CO$_2$ is a linear molecule with four fundamental vibrational modes. While the symmetric stretching mode $\nu_1$, is inactive in the IR, the bending mode $\nu_2$, which involves symmetric in-plane and out-of-plane bending, is active in the IR around 660 cm$^{-1}$ \citep{ehrenfreund1996, pontoppidan2008, suhasaria2017}, as is the asymmetric stretching mode $\nu_3$ near 2350 cm$^{-1}$ \citep{ehrenfreund1996, gerakines1999}. Furthermore, two combination bands ($\nu_1 + \nu_3$) around 3708 cm$^{-1}$ and ($2\nu_2 + \nu_3$) near 3600 cm$^{-1}$ \citep{gerakines1999} are also observed in the mid-infrared. The $^{13}$CO$_2$ isotope shows a small absorption peak around 2280 cm$^{-1}$ \citep{ehrenfreund1997}.

In addition to vibrational modes, CO$_2$ optical phonon modes due to lattice vibrations can also appear in mid-infrared spectra under certain conditions. In the laboratory with transmittance infrared spectroscopy at normal incidence, only transverse optical (TO) phonons are excited, seen as the typical $\nu_2$ and $\nu_3$ modes. However, when transmittance infrared spectroscopy is performed at oblique incidence \citep{baratta2000, cooke2016}, or in reflection absorption infrared spectroscopy (RAIRS) with light polarized parallel to the plane of incidence, both the $\nu_2$ and $\nu_3$ modes split, generating additional longitudinal optical (LO) modes. This phenomenon occurs only when the film thickness on the surface is smaller than the wavelength of the incident radiation \citep{berreman1963}.

For this reason, astronomical infrared spectra are usually compared with laboratory transmission spectra obtained at normal incidence. However, interstellar alignment arising from asymmetric dust grains could produce polarized radiation in star forming regions, potentially generating optical phonon modes \citep{andersson2015}. In such scenarios, comparisons with RAIRS or oblique-incidence transmittance spectra become necessary. Although, this comparison could still be challenging due to the presence of these additional phonon modes. Nevertheless, RAIRS is preferred to transmission infrared spectroscopy for studying physical changes in thin film of ice layers due to its higher sensitivity.

Although many previous laboratory studies have examined binary mixtures of CO$_2$ ice with common interstellar ice components \citep[e.g.][]{ehrenfreund1997, ehrenfreund1999, cooke2016} there are fewer studies on CO$_2$ layers sequentially deposited above or below these components \citep{van2006, edridge2013}. These studies have been invaluable for interpreting interstellar ice observations, but the influence of the underlying grain surface on the infrared profile remains unexplored. The study of layered CO$_2$ ices is also important because CO$_2$ prefers to cluster with itself and segregate from ice mixtures rather than remain mixed with other components when heated above 50 K \citep{oberg2009}, effectively forming a layered structure.

In this work, we explore three key effects on the CO$_2$ infrared profile: (1) the influence of an amorphous MgFeSiO$_4$ dust grain analog layer, (2) the impact of polar and apolar ice layers on top of the silicate surface, and (3) the changes observed during the warming of both ice layered systems until CO$_2$ desorption. The infrared data were then qualitatively compared with observational data from the James Webb Space Telescope obtained along the lines of sight of young and embedded protostars, particularly in relation to the $^{13}$CO$_2$ profile.

\par
\noindent
\par

\label{sec:intro}

\begin{figure*}[t]
\centering
\includegraphics[width=0.85\textwidth]{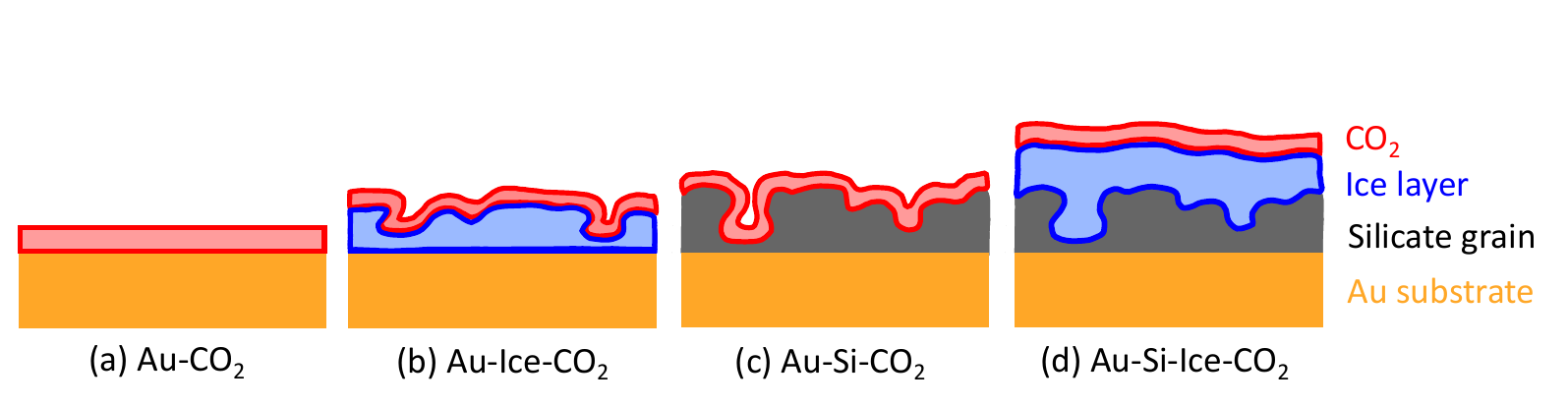}
\caption{Schematic of the layered systems: CO$_2$ deposited on (a) bare gold (Au), (b) water ice on gold, (c) a silicate film on gold, and (d) ice on a silicate film on top of gold. In panel (b), the ice layers consist of different thicknesses of pASW, while in panel (d), the ice layers are either apolar (CO, CH$_4$) or polar (pASW, CW, CH$_3$OH, or CH$_3$OH:H$_2$O in a 1:1 ratio).}
\label{fig:slab}
\end{figure*}

\section{Experimental} \label{sec:exp}
\subsection{Setup} \label{subsec:expset}
The setup consists of an ultra-high vacuum (UHV) chamber and two triply differentially pumped and highly collimated molecular beam lines, which has been described in detail in prior works \citep{he2016_,he2018b}. Here we briefly summarize the main features of the setup that are most relevant for the present study. The UHV chamber is a 10 inch diameter stainless steel chamber which is pumped by a combination of scroll pumps, turbomolecular pumps, and a cryopump to a base pressure of $2 \times 10^{-10}$ mbar after bake-out.

A bare polished gold-plated copper disk and a thin film of amorphous MgFeSiO$_4$ grains deposited on its surface were used as samples. The dust film on the disk was prepared by pulsed laser ablation of a rotating Mg-Fe silicate target in a quenching gas atmosphere of He/O$_2$ with a ratio of 6:1 at 6 Torr pressure. The exact method including the extraction and condensation of the grains onto the disk is described in detail in \citet{sabri2013}. The prepared samples were characterized by high-resolution transmission electron microscopy to determine the composition and internal structure as well as particle sizes, and by field-emission electron microscopy to determine the general morphology. Individual amorphous silicate grains with diameters between 4 and 8 nm forming chain-like aggregates were confirmed. The porosity of the films was found to be close to 90\%, leading to a surface-area-to-mass ratio of $\sim$1000 m$^2$ g$^{-1}$. A quartz crystal microbalance was used to monitor the thickness of $\sim$100nm of film on the gold plated-copper disk.

The substrate has been changed between different sets of experiments. First, the gold plated copper disk was clamped to the cold finger of the closed-cycle helium cryostat, followed by the silicate grain coated disk. The samples can be cooled down to 10 K and locally heated up to 300 K using a 50 W resistive heater situated right behind the sample holder. The temperature was measured using a calibrated silicon diode sensor and controlled by a Lakeshore 336 temperature controller to an accuracy of 0.05 K. A thermal radiation shield surrounds the sample and insulates it by reducing the radiative heat transfer from the surrounding. Carbon monoxide (CO) (99.97 per cent; Westfalen AG), methane (CH$_4$) (99.995 per cent; Westfalen AG), carbon dioxide (CO$_2$) (99.995 per cent; Westfalen AG), water (H$_2$O) and methanol (CH$_3$OH) (Emparta ACS) were deposited on the samples from the background through two UHV variable leak valves (VAT variable leak valve 590), one of which was reserved for liquids and the other for gasses. Although the gas was used without further purification, the liquids underwent three freeze–pump–thaw cycles to remove dissolved air. The CH$_3$OH:H$_2$O (1:1) mixture was prepared in a separate mixing glass bulb by adjusting the vapor pressure ratio of the purified liquids. The mixing ratio was verified by measuring the quadrupole mass spectrometer signal during deposition. The molecules were deposited on the samples at 15 K. The chamber is equipped with a Fourier transform infrared (FTIR) spectrometer (Bruker Invenio R) and a MCT detector for the analysis. Infrared spectra were collected after ice deposition and also during warming-up experiments with a linear ramp rate of 1 K min$^{-1}$ across a temperature range of 15 to 160 K. These measurements were taken in the RAIRS configuration, with a grazing incidence angle of 78$^{\circ}$, covering the spectral range from 5000 to 800 cm$^{-1}$ with a resolution of 1 cm$^{-1}$. After deposition, 128 scans were recorded, while during warming-up scans were taken at every 5 K.

\subsection{Methods}
\subsubsection{Procedure}
We conducted three sets of experiments. The differently layered systems that were used are visualized in \autoref{fig:slab}. In the first set, we examined the effect that (c) a thin amorphous silicate layer deposited on a gold surface has on the RAIRS spectra of pure CO$_2$. The results were compared to measurements performed on (a) a bare gold. Furthermore, we investigated the impact of (b) varying water ice thicknesses on the RAIRS spectra of CO$_2$ compared to both the gold and silicate surfaces.

In the second set, we studied how the infrared profile of CO$_2$ changes when present on top of (d) pure and mixed polar and apolar ice layers. For pure ices, we selected common species such as H$_2$O, CO, CH$_4$, and CH$_3$OH, while a CH$_3$OH:H$_2$O (1:1) was used as the surface for the mixed ice. In all cases, pure or mixed ice layers were deposited first, followed by CO$_2$ ice. At 15 K, the water ice layer exhibited an amorphous structure and was highly porous, hence denoted porous amorphous solid water (pASW). We also performed experiments on crystalline water (CW) by depositing water at 150 K, holding it at that temperature, and then cooling it back to 15 K. We used two different coverages of CO$_2$ in the ice layers for comparison.

In the final set of experiments, we conducted warming experiments on the ice layers visualized in panel (d) of \autoref{fig:slab} using the higher CO$_2$ coverages from the second set to ensure a high signal-to-noise ratio across all peaks. These measurements were intended to determine the temperatures at which CO$_2$ molecules undergo orientational changes, crystallization, and diffusion, as well as the onset of desorption.

\subsubsection{Coverage determination}
The molecular column density ($\text{molecules cm}^{-2}$) of the deposited ices is calculated from a modified Beer–Lambert Equation (\ref{BL}) to account for the RAIRS mode configuration \citep{ioppolo2022}.

\begin{equation}
    C=ln(10)\frac{sin(\theta)}{2A}\int \tau(\nu) d\nu \label{BL}
\end{equation}

where $A(\nu)$ is the band strength (in cm molecule$^{-1}$), $\tau(\nu)$ is the optical depth and $\theta$ = 12$^{\circ}$ is the angle of incidence of the IR light with respect to the substrate. The column density was then expressed in terms of a monolayer (ML, around $10^{15}$ molecules cm$^{-2}$). The band strengths are taken from \citet{bouilloud2015}, with \autoref{tab:rangesused} summarizing the values and integration ranges used. 
\vspace{-0.5cm}

\begin{deluxetable*}{cccc}
\tablenum{1}
\tablewidth{0pt}
\tablehead{
\colhead{species} & \colhead{band} & \colhead{integration range [cm$^{-1}$]} & \colhead{band strength [cm molecule$^{-1}$]}}
\startdata
         CO$_2$ & asymmetric $^{12}$CO$_2$ stretch $\nu_3$ & 2379-2332 & $6.8\cdot10^{-17}$ \\
         CO & $^{12}$CO stretch $\nu_1$ & 2160-2127 & $1.12\cdot10^{-17}$\\
         CH$_4$ & C-H stretch $\nu_3$ & 3029-2990 & $1.1\cdot10^{-17}$ \\
         H$_2$O & O-H stretch $\nu_3$ & 3672-3060 & $1.5\cdot10^{-16}$\\
         CH$_3$OH & C-O stretch $\nu_8$ & 1065-1018 & $1.07\cdot10^{-17}$ \\
         \enddata
       \caption{The values for the band strength are taken from \citet{bouilloud2015}.}\label{tab:rangesused}
\end{deluxetable*}
\vspace{-0.5cm}
\section{Results and Analysis} \label{sec:exp}
\subsection{Effect of silicate grain on CO$_2$ IR profiles}

\begin{figure*}[!ht]
\vspace{-1.25cm}
\includegraphics[width=0.5\textwidth]{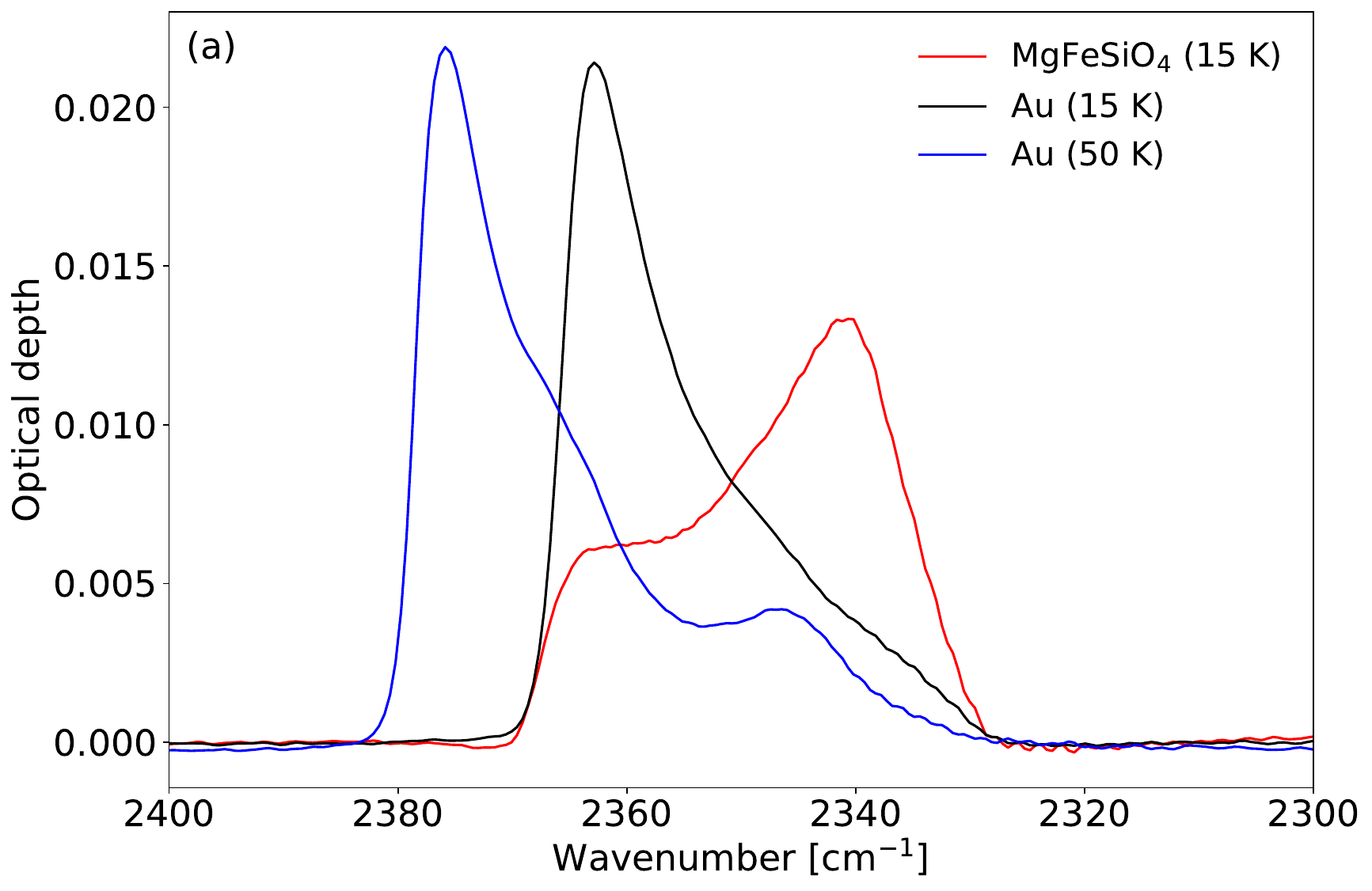}
\includegraphics[width=0.5\textwidth]{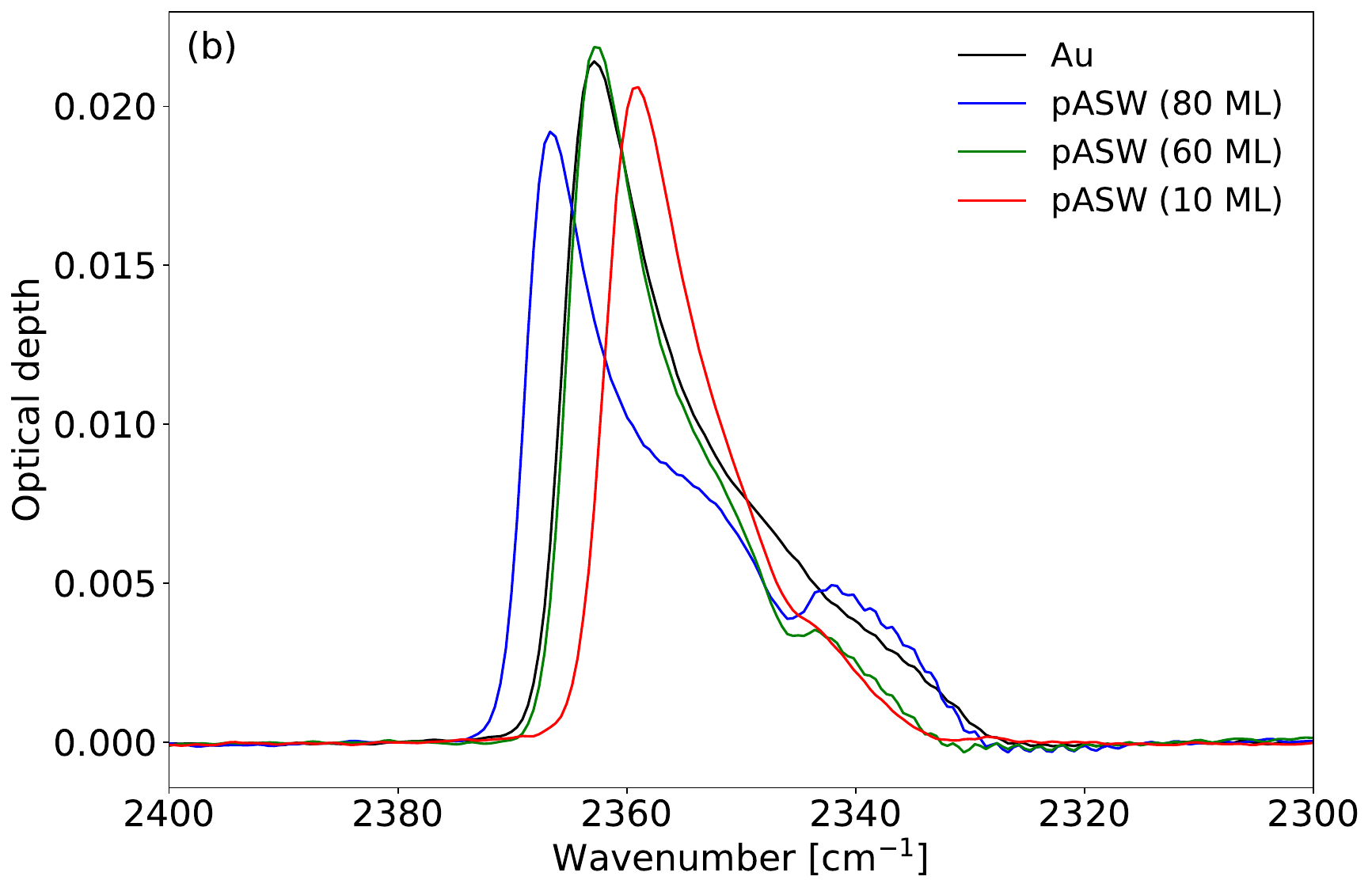}
\caption{(a) RAIRS spectra of $\sim$1 ML CO$_2$ ($\nu_3$ vibrational mode) deposited on bare gold at 15 and 50 K and on gold coated with a 100 nm amorphous silicate film at 15 K. (b) Spectra of $\sim$1 ML CO$_2$ deposited on varying thickness of porous amorphous water layers on gold at 15 K are shown alongside the spectrum on bare gold for comparison.}\label{fig:LOTOsplitting}
\end{figure*}

We deposited $\sim$1 ML of CO$_2$ on the gold surface at 15 K as seen in \autoref{fig:LOTOsplitting} (a). We observed only the LO mode at 2363 cm$^{-1}$, indicative of a disordered amorphous state. According to the metal surface selection rule (MSSR), only the perpendicular components of molecular vibrations are infrared active and can be observed using RAIRS \citep{fulker2025}. This explains why, on a bare gold surface, only the LO component was detected. We also observed a broad red shoulder. Furthermore, we detected weak and broad signals for two combination bands: ($\nu_1 + \nu_3$) at 3706 cm$^{-1}$ and ($2\nu_2 + \nu_3$) at 3600 cm$^{-1}$, along with a $^{13}$CO$_2$ peak at 2280 cm$^{-1}$ (not shown in \autoref{fig:LOTOsplitting}). When CO$_2$ was deposited on the gold surface at 50 K, the LO mode blue shifted to 2376 cm$^{-1}$ and a small TO mode appeared at 2346 cm$^{-1}$. In addition to the TO mode, a weak shoulder appeared at 2365 cm$^{-1}$ that replaced the broad red shoulder observed at 10 K. The two combination modes became sharper, but their band positions remained nearly the same, while the $^{13}$CO$_2$ peak blue shifted to 2283 cm$^{-1}$, probably due to the fact that CO$_2$ ice is already crystalline. The appearance of a small TO mode for CO$_2$ deposited at 50 K on a bare gold surface was not surprising, as it has already been seen when 30 ML CO$_2$ ice deposited on gold at 10 K was heated to around 60 K \citep{he2018b}. This indicates that MSSR could be influenced by the ice phase.

In another experiment, when a similar amount of CO$_2$ was deposited on a silicate-covered gold surface at 15 K, the LO mode at 2363 cm$^{-1}$ was significantly reduced, while the TO mode was enhanced and appeared at 2341 cm$^{-1}$. This suggests that the MSSR was relaxed on the grain film, despite the presence of the underlying metal layer. A similar behavior was previously noted when N$_2$O films were deposited on copper covered with amorphous silica, with a more pronounced effect for a silica film thickness of around 300 nm \citep{lasne2015}. In our case, the effect was observed with a silicate film of only 100 nm thickness. The comparison is relevant because the stretching mode band strength of N$_2$O \citep{fulvio2009} is very similar to that of CO$_2$. However, it is also possible that the grain composition, in addition to thickness, influences MSSR. The combination modes and $^{13}$CO$_2$ feature remained unaffected by the presence of silicates. At 50 K, the TO mode remained unchanged and the LO mode is blue shifted to 2376 cm$^{-1}$, similar to the behavior observed on gold. Other modes followed the same pattern as CO$_2$ deposited on gold at 50 K, hence they are not shown.

\subsection{Impact of amorphous water layer with varying thickness on CO$_2$ IR profiles on a bare gold surface}
To check whether this relaxation of MSSR could be achieved without the silicate layer, water ice of three different thicknesses was deposited directly on gold. We then compared the infrared behavior of $\sim$1 ML CO$_2$ on these water ice layers to the same CO$_2$ thickness on gold, as shown in \autoref{fig:LOTOsplitting} (b). With a 10 ML thick water ice layer, the CO$_2$ infrared spectrum closely resembled that on bare gold, except that the LO peak was red-shifted to 2359 cm$^{-1}$. When 60 ML of water ice was deposited, the behavior remained similar to gold, especially with regard to the peak position, but a small TO peak appeared at 2343 cm$^{-1}$. At 80 ML thickness, the TO peak became stronger, and the LO peak blue-shifted to 2367 cm$^{-1}$; however, unlike with silicates, the MSSR was not fully relaxed. This emphasizes the continued influence of the underlying grain surface on the infrared behavior.

RAIRS measurements rely on the reflective properties of metal surfaces, which do not accurately represent the composition of cosmic dust grains mainly made of carbonaceous and siliceous materials that are primarily amorphous \citep{henning2010}. Additionally, the absorption profiles of the CO$_2$ stretching mode are influenced by both the polarization and the incidence angle of the radiation. As a result, direct comparisons between observational data and laboratory measurements in RAIRS are challenging. Our results indicate that the enhancement of the LO mode is reduced when the metal surface is coated with a layer of amorphous silicate material, making RAIRS spectra more comparable to transmission spectra that only show peaks at the position of the typical TO mode. Based on this, we propose that laboratory RAIRS studies on grain coated or grain and ice layers on metal surfaces provide more reliable references for observational data, keeping the advantage of RAIRS intact, as these spectra have a higher sensitivity at the sub-monolayer/monolayer coverage regime. This approach enables RAIRS spectra to be used for qualitative comparisons, making the technique more versatile.

\subsection{Effect of polar/apolar ices on top of the silicates on CO$_2$ IR profiles}
\begin{figure*}[!ht]
\resizebox{\hsize}{!}{\includegraphics{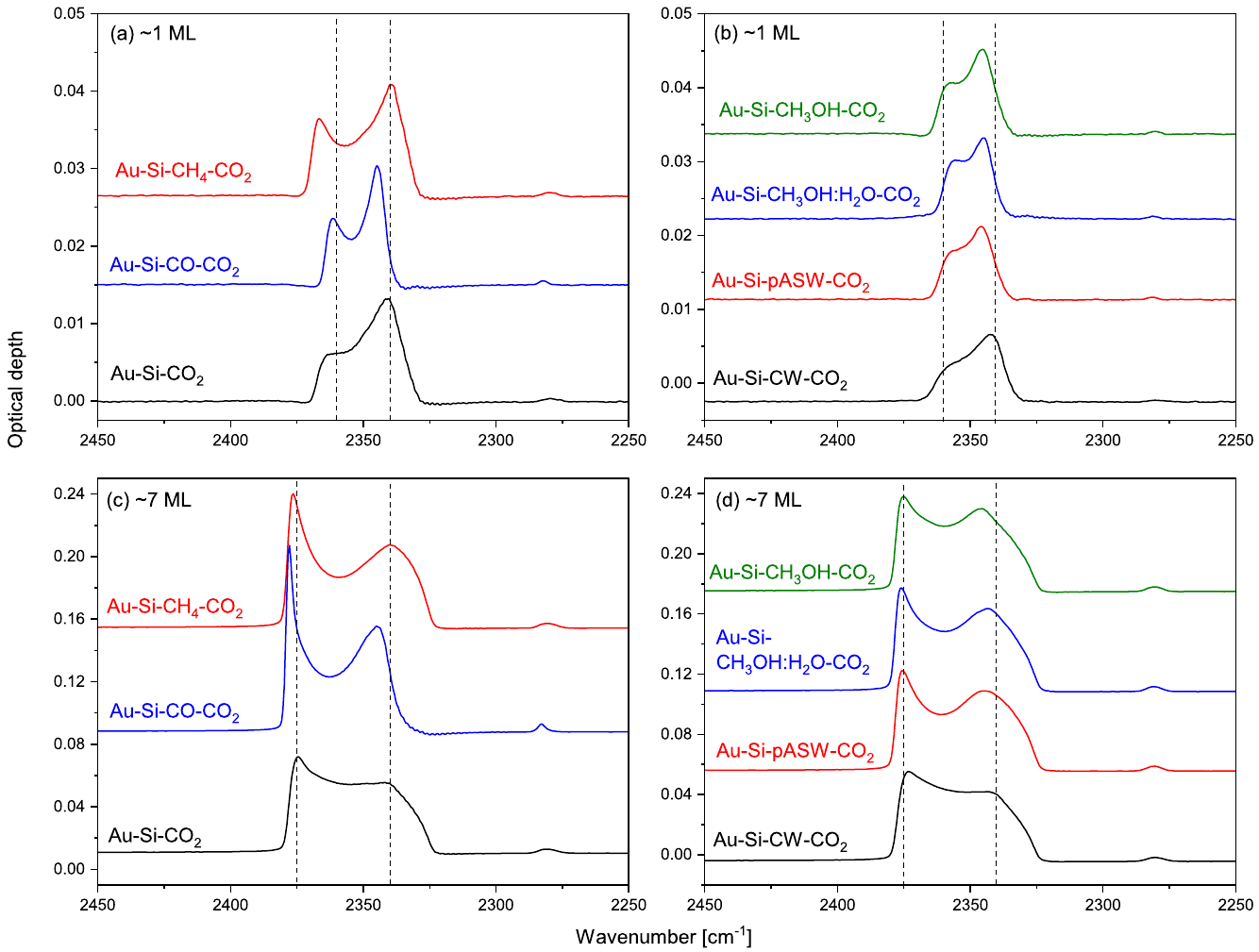}}
\caption{RAIRS spectra of CO$_2$ ($\nu_3$ vibrational mode) deposited at 15 K on gold coated with a 100 nm amorphous silicate film, with or without an additional $\sim$10 ML layer of pure or mixed ice. The top column shows 1 ML CO$_2$ spectra on (a) the silicate film without additional ice layer and on pure apolar ices, (b) on pure polar ices and an ice mixture. The bottom panels (c) and (d)} display the corresponding spectra for 7 ML of CO$_2$. Spectra have been vertically offset for clarity. Vertical lines at 2375, 2360 and 2340 cm$^{-1}$, provides a visual reference. The first two values (2375 and 2360 cm$^{-1}$) are typically close to the LO mode, while the latter value (2340 cm$^{-1}$) to the TO mode.\label{fig:(a)polar}
\end{figure*}

{\autoref{fig:(a)polar} shows the RAIRS spectra of CO$_2$ deposited on apolar ices (CO, CH$_4$), polar ices (pASW, CW, and CH$_3$OH), a polar ice mixture (CH$_3$OH:H$_2$O), and directly on silicate grains (MgFeSiO$_4$) without an additional ice layer. To investigate the effect on the $^{13}$CO$_2$ peak and the combination modes, additional RAIRS spectra were also recorded for 7 ML of CO$_2$ deposited on silicate grains, with and without an additional ice layer. When an additional layer of ice was deposited on silicates, the TO mode still appeared stronger than the LO mode for the case where 1 ML of CO$_2$ was added. However, for a higher CO$_2$ coverage, the LO mode increases in intensity and becomes nearly equal in intensity to the TO mode. Depending on the additional layer, we see differences in the band positions for both LO and TO modes of CO$_2$, clearly showing the effect of the chemical environment. 

A clearer separation between the LO and TO vibrational modes and a deeper trough is observed when 1 ML CO$_2$ is deposited on apolar ices, compared to polar ices as evident in {\autoref{fig:(a)polar} (a) and (b). For CO$_2$ on CH$_4$, CW, or directly on silicates, the TO mode appears closer to 2340 cm$^{-1}$, whereas in all other cases it appears near 2345 cm$^{-1}$. The LO mode is blue-shifted to 2367 cm$^{-1}$ for CO$_2$ on CH$_4$ and to 2363 cm$^{-1}$ on silicates, while for the remaining cases, it appears close to 2360 cm$^{-1}$. The CO$_2$ peaks on CO appear slightly sharper than on other ice layers which becomes clear when a higher coverage (7 ML) of CO$_2$ is added. 

In the case of 7 ML CO$_2$ deposited on a CH$_4$ layer, the TO mode appears at nearly the same position as on silicate, while the LO mode is slightly blue-shifted to 2376 cm$^{-1}$ as evident in {\autoref{fig:(a)polar} (c). In contrast, for CO$_2$ deposited on a CO layer, both the LO and TO modes are blue-shifted to 2378 and 2345 cm$^{-1}$, respectively, with the LO mode appearing sharper compared to those on silicate or CH$_4$-covered silicate. This sharpening and blue shift can be attributed to the CO$_2$ being deposited on pre-existing polycrystalline CO ice \citep{he2021}. The $^{13}$CO$_2$ feature at 2280 cm$^{-1}$ and the combination modes at 3705 and 3598 cm$^{-1}$ remain similar for CH$_4$ on silicate and bare silicate. However, for CO on silicate, all features are sharper and blue-shifted to 2283, 3709, and 3600 cm$^{-1}$, respectively. 

When CO$_2$ is deposited onto these polar ices or mixtures, the peak shapes and positions are not significantly affected as seen in {\autoref{fig:(a)polar} (d). CO$_2$ deposited on silicate and crystalline water (CW) displays notable similarity with minimal separation between the LO and TO modes, despite the underlying layer being fully crystalline in the case of CW. In contrast to CO$_2$ on polycrystalline CO, the crystalline nature of CW does not sharpen the CO$_2$ features. The LO mode is blue-shifted by 1 cm$^{-1}$ for CW compared to silicate.

For CO$_2$ deposited on other polar ices or ice mixtures, the LO and TO modes are more clearly separated than on silicate. The TO mode appears at 2345, 2346, and 2343 cm$^{-1}$, while the LO mode is found at 2375, 2375, and 2376 cm$^{-1}$ for pASW, CH$_3$OH, and CH$_3$OH:H$_2$O, respectively. There is no noticeable shift in the $^{13}$CO$_2$ peak or the combination modes.

\begin{figure*}[ht]
\centering
\includegraphics[width=0.75\textwidth]{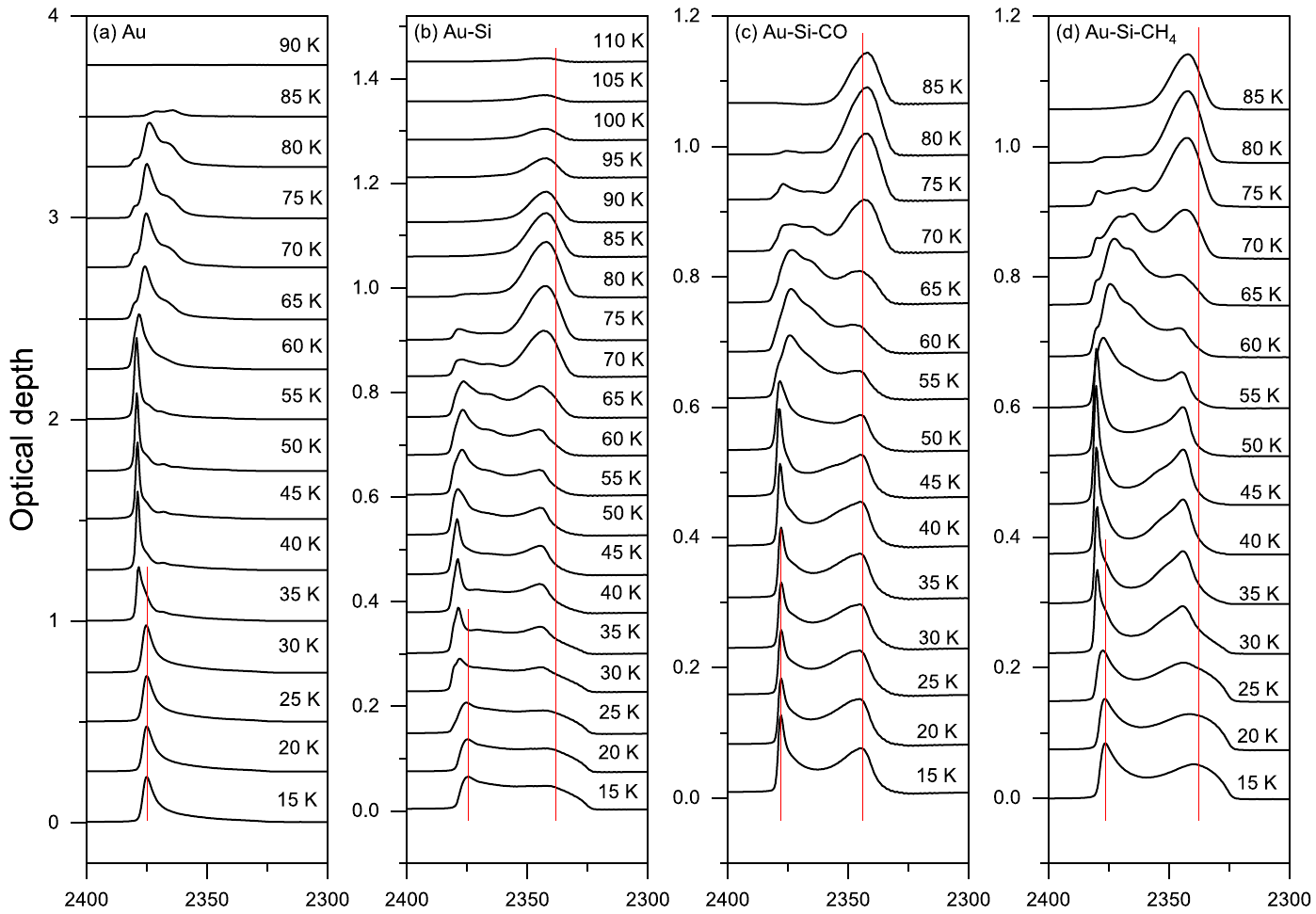}
\includegraphics[width=0.75\textwidth]{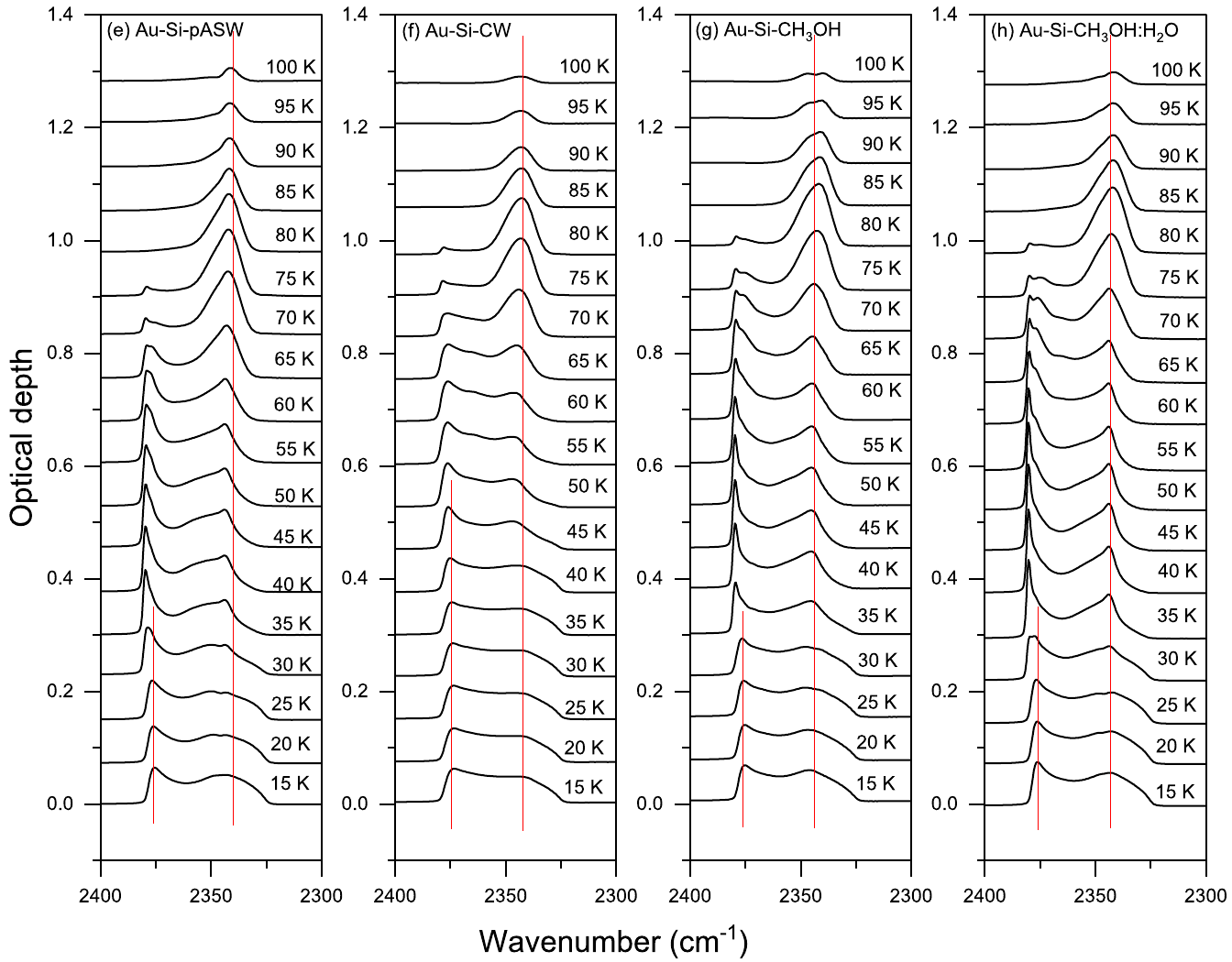}
\caption{RAIRS spectra of $\sim$7 ML CO$_2$ ($\nu_3$ vibrational mode) during heating from 15 K to 160 K. The CO$_2$ was deposited on bare gold, gold coated with a 100 nm amorphous silicate film, and on silicate films with an additional layer of pure or mixed ice ($\sim$10 ML). The temperatures at each stage of heating are labeled above each curve, with spectra for silicate/silicate-ice layers offset by 0.075 and those on gold offset by 0.25 for clarity. The top row displays CO$_2$ spectra on: (a) bare gold, (b) the silicate film, (c) CO on silicate film, and (d) CH$_4$ on silicate film. The bottom row presents spectra for ice layers on silicate films: (e) porous amorphous solid water (pASW), (f) crystalline water (CW), (g) pure CH$_3$OH, and (h) CH$_3$OH:H$_2$O mixture. The vertical lines at close to LO and TO position provides a visual reference.}
\label{fig:heating}
\end{figure*}
\subsection{Effect of heating on CO$_2$ IR profile}

As CO$_2$ ice is heated from 15 K, it starts to undergo a phase transition, rearranging from a disordered amorphous state to a micro-/polycrystalline structure. With further heating, the ice crystallizes fully at a specific temperature. As the temperature continues to rise, the CO$_2$ molecules begin to diffuse across the surface, and at sufficiently high temperatures, they desorb, which means that they leave the surface (see \citet{escribano2013} for an extensive discussion on the temperature effects on the CO$_2$ IR profile). How these four physical phenomena are affected by CO$_2$ present on a bare gold surface, a silicate-covered gold surface and 10 ML additional ice layer on a silicate covered gold surface is shown in \autoref{fig:heating}.

\begin{figure*}[!ht]
\includegraphics[width=0.33\textwidth]{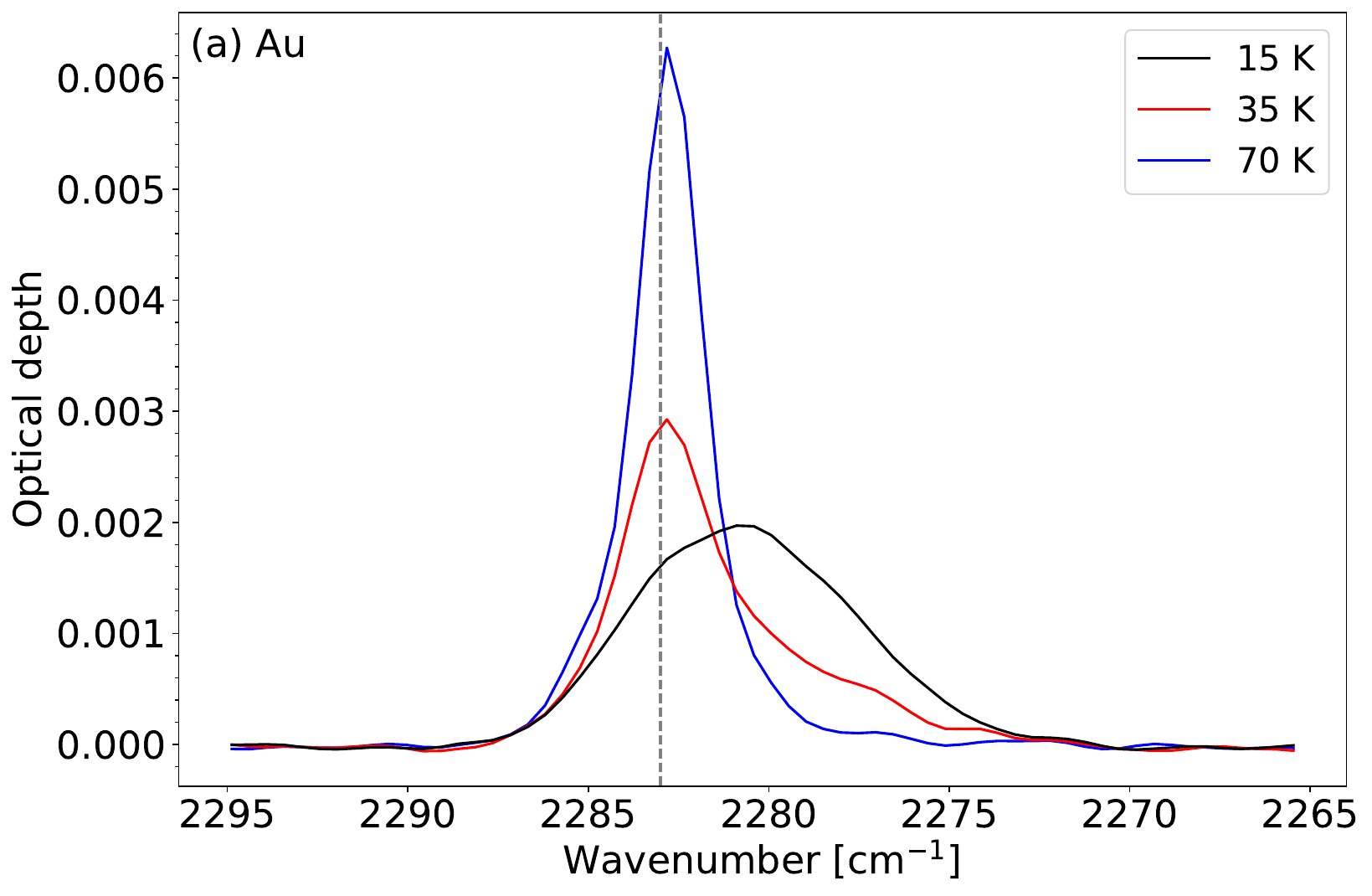}
\includegraphics[width=0.33\textwidth]{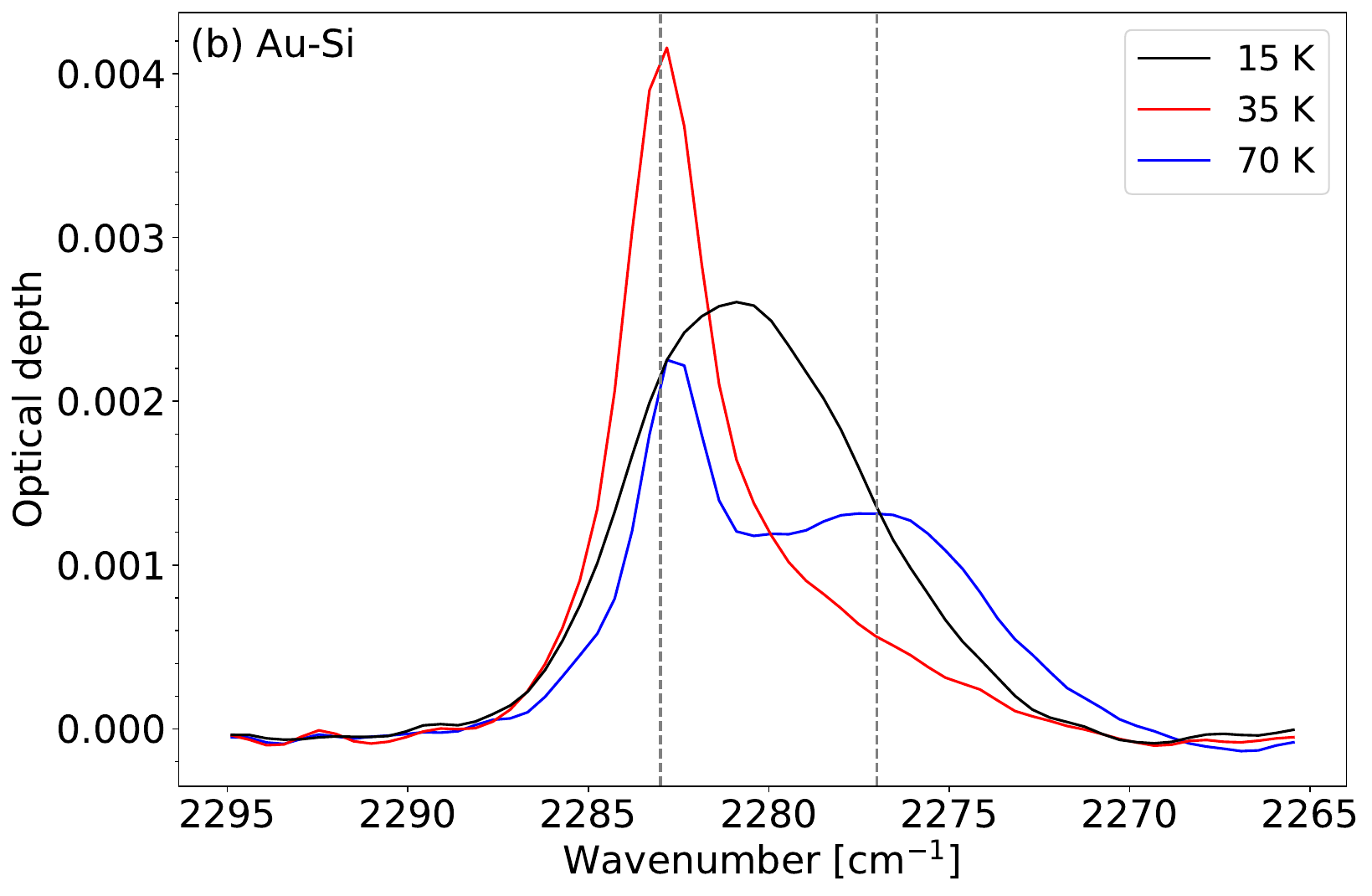}
\includegraphics[width=0.33\textwidth]{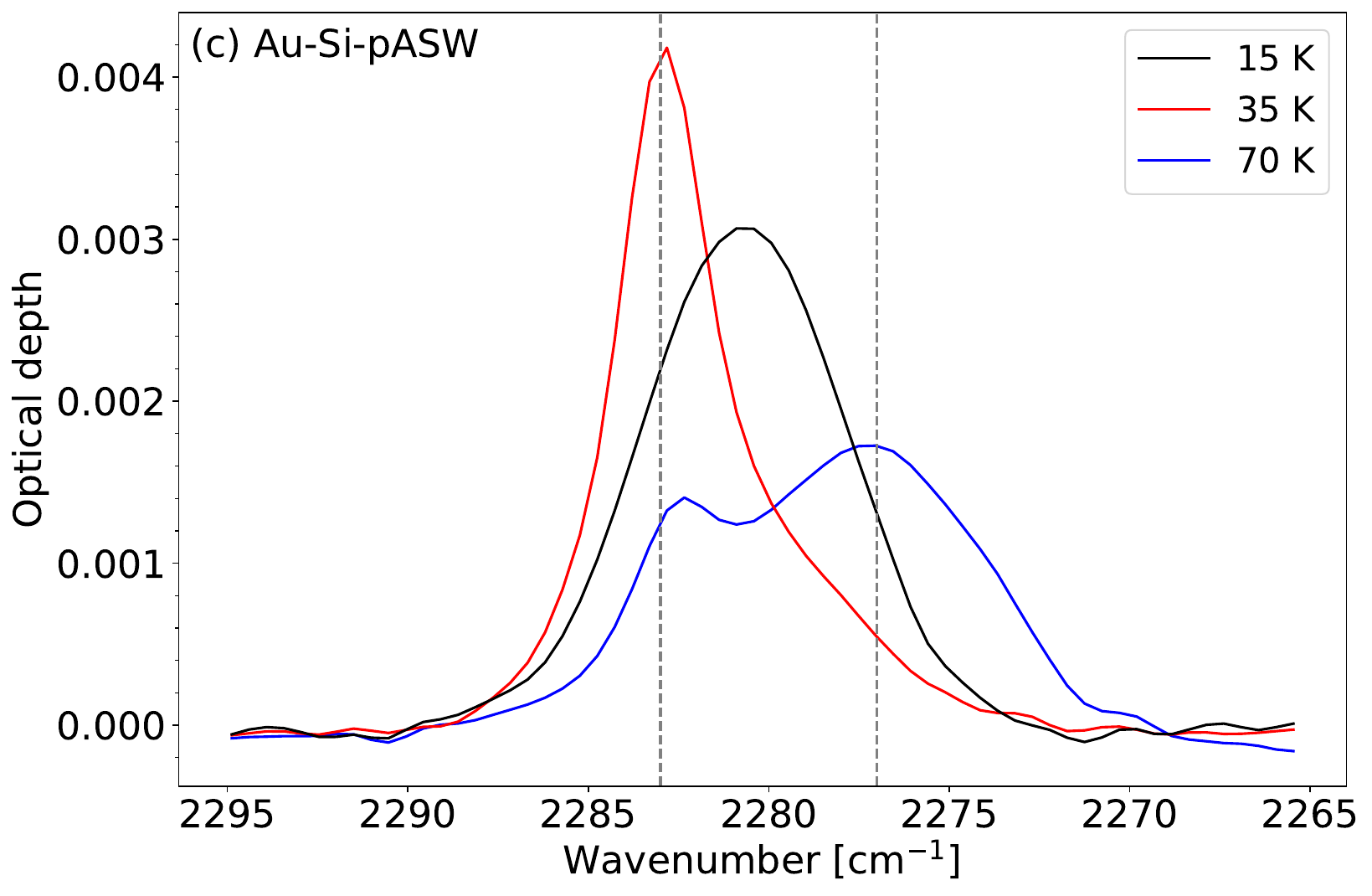}
\caption{RAIRS spectra of $\sim$7 ML CO$_2$ ($\nu_3$ $^{13}$CO$_2$ vibrational mode) shown at selected temperatures during warming up. The CO$_2$ was deposited on (a) bare gold, (b) gold coated with a 100 nm amorphous silicate film, and on (c) a silicate film with 10 ML pASW ice on top. The vertical lines at 2283 and 2277 cm$^{-1}$ provide a visual reference.}
\label{fig:13CO2}
\end{figure*}

\subsubsection{Phase transition}
During the phase transition, the $\nu_3$ bands for $^{12}$CO$_2$, $^{13}$CO$_2$, and the combination modes ($\nu_1$ + $\nu_3$) and (2$\nu_2$ + $\nu_3$) sharpen. The LO and TO modes blue shift to higher wavenumbers, $\sim$2379 cm$^{-1}$ and $\sim$2345 cm$^{-1}$, respectively, for all the cases. The $\nu_3$ band for $^{13}$CO$_2$ and the combination modes also blue shift by $\sim$2 cm$^{-1}$. For the $\nu_3$ band of $^{12}$CO$_2$, the combined effect of blue shift and the peak sharpening begins at 35 K for all cases, except for CO, CH$_4$, and CW. As mentioned earlier, CO ice is already polycrystalline at 15 K, so CO$_2$ deposited on CO ice is already in a more ordered state, and no changes occur from 15 K to above 40 K. For CH$_4$, the transition occurs between 25 and 30 K, since the underlying CH$_4$ surface also undergoes reorientation within this range \citep{emtiaz2022}. Although CW is fully crystalline, CO$_2$ ice deposited at 15 K does not initially orient toward a more ordered state, with the phase transition occurring later, around 45 K. A possible explanation is that on a surface with a mixed phase (partially amorphous and crystalline), nucleation for crystalline ice phase growth is already present. In contrast, on fully crystalline CW water, there is a lack of pre-existing nucleation sites.

\subsubsection{Complete crystallization}
As the CO$_2$ ices on different surface layers are heated beyond 35 K, all the bands become sharper and so no significant shifts occur. The combination mode serves as a more reliable complete crystallinity marker, as it exhibits the same behavior in both reflection and transmission modes \citep{he2018b}. When this mode is at its sharpest, it indicates nearly complete crystallization. In all cases, the sharpest combination mode occurs almost around 50 K, aligning well with the temperature range reported in previous studies \citep{falk1986, isokoski2013}.

\subsubsection{Diffusion}
For CO$_2$ deposited on top of the bare and ice-covered silicate grain, a new peak at ~2276 cm$^{-1}$ appears at 60 K as a red shoulder to the $\nu_3$ band of $^{13}$CO$_2$. As the temperature increases, it grows in intensity. By 75 K, it develops into a double peak as seen in panels (b) and (c) of \autoref{fig:13CO2}. The growth of the double peak feature cannot be observed for CO$_2$ deposited on gold without the silicate and is hence not visible in panel (a). Previous studies show that CO$_2$ diffusion on non-porous ASW becomes active above 60 K \citep{he2023}, allowing molecules to either form island clusters on the surface or penetrate through the bulk of water ice surface, both of which can affect the infrared peak profiles. A clearer effect on the $\nu_3$ band of $^{12}$CO$_2$ may require investigation with sub-monolayer coverages \citep{he2017}. Previous studies with pure CO$_2$ showed no significant changes around 60 K for the $\nu_3$ band of $^{13}$CO$_2$ \citep{he2018b}, but the underlying silicate layer in this experiment influenced the outcome. The initial red-shoulder peak serves as a marker for the onset of diffusion. On the other hand, the combination modes remain largely unaffected.

\subsubsection{Additional features during the warming up process}

For the $\nu_3$ band of $^{12}$CO$_2$, three additional changes in the IR peak profile were observed during warming up. First, LO mode intensity variations occur between 65 and 70 K, where the LO mode intensity decreases and subsequently the TO mode intensity increases. On gold, which only exhibits the LO mode, no decrease is observed until desorption begins. Second, starting at 60 K, the sharp LO mode undergoes a red shift for CO$_2$ on gold as well as on the CO and CH$_4$ ices, the latter two exhibiting larger shifts, probably due to their desorption. No shifts are observed for CO$_2$ on CW, CH$_3$OH, or CH$_3$OH:H$_2$O, while a small blue shift appears at 70 K in the case of the bare silicate substrate, although its cause remains unclear. Third, shoulder peaks emerge between the LO and TO mode at different temperatures and band positions depending on the surface. On gold, they appear at 40 K at 2375 and 2368 cm$^{-1}$ and shift to 2380 and 2366 cm$^{-1}$ at 65 K. On silicates, they arise at 35 K at 2371 cm$^{-1}$ and shift to 2365 cm$^{-1}$ at 55 K. For CO, the shoulder emerges at 60 K at 2365 cm$^{-1}$, while for CH$_4$, it appears at 30 K at 2376 cm$^{-1}$ and persists until 50 K, with new shoulders at 2380 and 2371 cm$^{-1}$ forming at 60 K. On pASW, the shoulder remains between 65 and 75 K at 2376 cm$^{-1}$, while on CW, it is present between 55 and 65 K at 2364 cm$^{-1}$. For CH$_3$OH, shoulders emerge at 2376 and 2371 cm$^{-1}$ at 35 K and persist until 80 K, whereas for CH$_3$OH:H$_2$O, they appear at 2376 cm$^{-1}$ at 50 K and remain until 80 K. In all cases, the shoulder peaks persist until the LO mode intensity declines.

\citet{edridge2013} observed a shoulder peak at 2373 cm$^{-1}$ for CO$_2$ adsorbed on graphite at 28 K, attributing it to a heterogeneous and amorphous CO$_2$ layer that disrupts phonon vibrations and modifies IR interactions. Upon heating to 55 K, this peak shifted to 2367 cm$^{-1}$. Additionally, a feature at 2361 cm$^{-1}$ was detected in a 15\% CH$_3$OH:H$_2$O ice mixture at 30 K, likely due to interactions between CO$_2$ and CH$_3$OH \citep{edridge2013}. Similarly, a peak at 2365 cm$^{-1}$ was identified when a CO$_2$ and H$_2$O mixture was deposited between 50 and 80 K, attributed to CO$_2$ trapped within partially ordered amorphous H$_2$O ice \citep{sandford1990}. Furthermore, DFT calculations have shown that vibrational bands from clusters of up to five CO$_2$ molecules in water can contribute to a peak at 2367 cm$^{-1}$ \citep{tychengulova2024}. The present study confirms the complexity of these shoulder peaks, which arise from dipole-dipole interactions between CO$_2$ and surrounding species, spanning from amorphous to crystalline forms. Importantly, this shoulder peak should not be mistaken for the X mode, which is described by \citet{escribano2013} as a low frequency shoulder to the $\nu_3$ band of $^{12}$CO$_2$ at 2328 cm$^{-1}$.

\subsubsection{Desorption}
Starting at 85 K, the infrared features of CO$_2$ on gold show a significant decrease in intensity, and beyond 90 K, the IR signal of CO$_2$ disappears completely. This marks the desorption temperature of CO$_2$ on the gold surface. In contrast, for CO$_2$ on silicate-coated gold, only the TO mode is observed at 85 K. Desorption begins beyond this temperature and completes between 115 and 120 K, which is significantly higher than on the gold surface. The porous nature and the likely presence of various binding sites on the silicate film cause CO$_2$ to remain bound to the surface much longer than on gold. The desorption temperatures for ice layers on silicate films were not determined, as all surfaces, except for the CW surface, undergo changes during the warming process. The volatile species CO and CH$_4$ are expected to desorb before CO$_2$, while pASW, CH$_3$OH, and CH$_3$OH:H$_2$O could trap some CO$_2$ in their pores and release it at higher temperatures. The desorption of these species from the same silicate grain has been studied using temperature programmed desorption experiments, the results of which will be published in an upcoming manuscript. Based on this study it can be said that most of CO and CH$_4$ desorbs around 70 K.

\begin{figure*}
\begin{center}
\resizebox{\hsize}{!}{\includegraphics{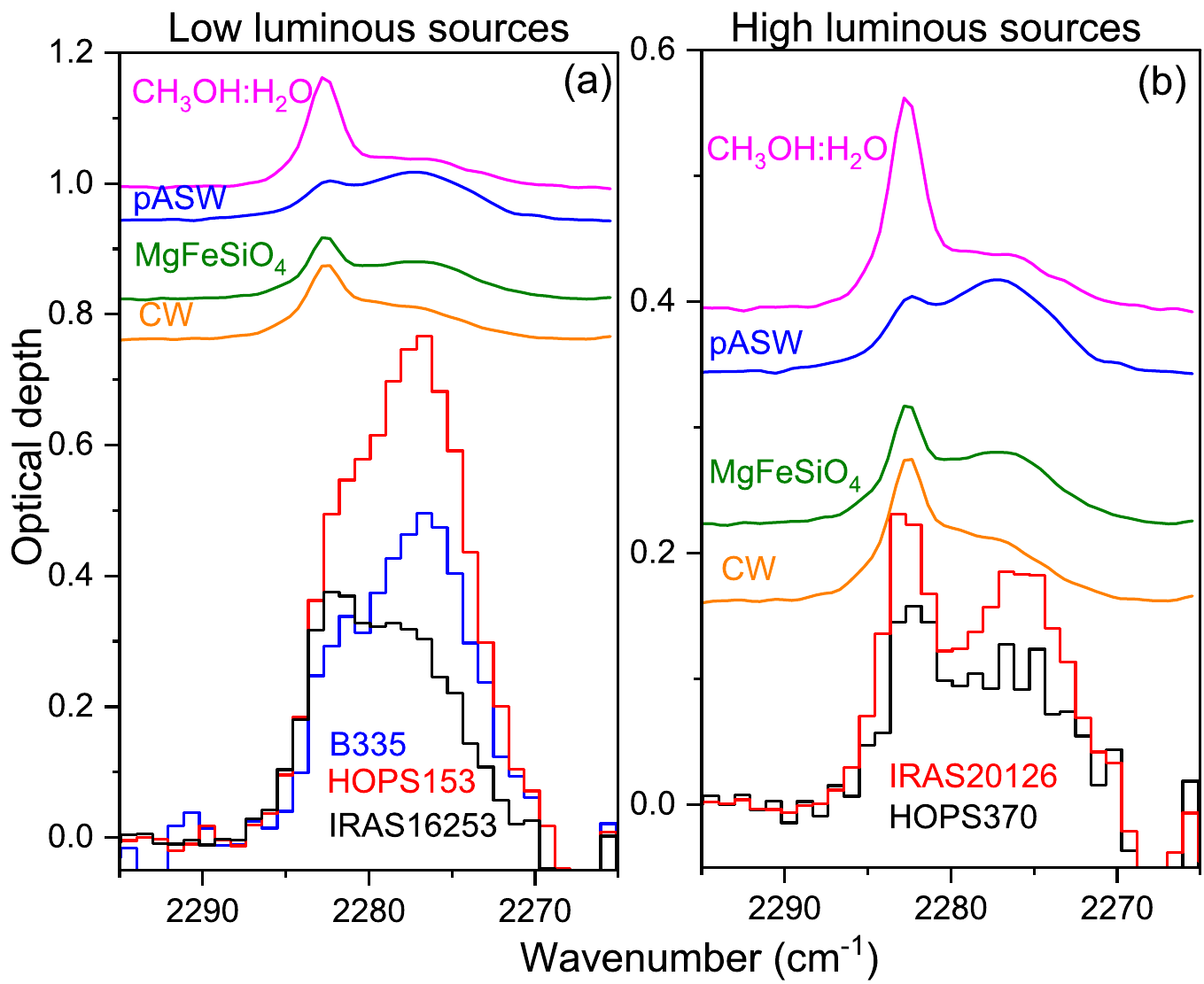}}
\caption{Comparison of the $^{13}$CO$_2$ ice feature toward (a) low luminous and (b) high luminous protostars with laboratory spectra on different surfaces. To facilitate comparison, all laboratory spectra have been scaled by a factor of 40. Additionally, in panel (b), the CH$_3$OH:H$_2$O spectrum is vertically offset by 0.4, while in panel (a), all laboratory spectra from panel (b) are offset by 0.6.}
   \label{fig5}
\end{center}
\vspace{-0.5cm}
\end{figure*}

\section{Astrophysical implications}

There are two important implications of our work. First, we observed a significant increase in the desorption temperature of CO$_2$ on our dust grain analogue, an amorphous silicate surface, compared to gold. About 50\% of CO$_2$ remained adsorbed on the silicate surface until a temperature of 88 K, while at that temperature CO$_2$ on the gold surface was nearly fully desorbed. This suggests that CO$_2$ will return to the gas phase much later during the warm-up phase of star formation in silicate grain environments. Consequently, more CO$_2$ molecules will remain locked in the solid state, increasing the likelihood of their active participation in grain surface chemistry for the formation of complex organic molecules. The higher binding energy of CO$_2$ on silicate grains will also shift the CO$_2$ snowline in protoplanetary disks to higher temperatures, which could influence the planet formation process \citep{minissale2022}.

Second, our laboratory infrared spectra of $^{13}$CO$_2$ on silicates and silicates covered by additional ice layers pure or in mixtures revealed a distinct double peak profile at 70 K that was absent on the gold surface. In particular, this double peak structure resembles recent JWST observations of $^{13}$CO$_2$ toward the young, embedded protostars HOPS 370 and IRAS 20216 \citep{brunken2024}. The observed $^{13}$CO$_2$ spectral profiles were analyzed in the past by fitting them with combinations of laboratory $^{13}$CO$_2$ spectra obtained at different temperatures and for different ice mixtures. The best fit was achieved using a combination of three different profiles, which are warm CO$_2$:CH$_3$OH at 105 K, hot CO$_2$:H$_2$O at 160 K, and pure CO$_2$ at 80 K (see Figure 5 in \citet{brunken2024}). However, none of the individual profiles exhibited a double-peak profile.

For completeness, previous laboratory measurements of $^{13}$CO$_2$ infrared spectra have shown a distinct double-peaked profile in transmission only above 110 K, when CO$_2$ was mixed with H$_2$O and CH$_3$OH. In these cases, the double peak was attributed to the segregation of CO$_2$ from the ice mixture \citep{boogert2000}. These laboratory $^{13}$CO$_2$ profiles were then used also for interpretation of ISO-SWS observations along 13 galactic lines of sight, including high- and low-mass protostars and molecular clouds \citep{boogert2000}.


To compare our laboratory spectra with observations, we use archival JWST data obtained with the Near-Infrared Spectrograph (NIRSpec) from the Investigating Protostellar Accretion (IPA) program \citep{federman2024,brunken2024}. Following the analysis by \citet{brunken2024}, we extract the NIRSpec spectrum, $F_\mathrm{obs}$, from the source positions \citep[listed in Table A.1 in][]{brunken2024} with an aperture radius of 0.6$''$. We estimate the local continuum, $F_\mathrm{cont}$, by fitting a cubic polynomial considering both sides of the $^{13}$CO$_2$ absorption feature at 4.4\,$\mu$m ($\approx$2280\,cm$^{-1}$). The optical depth spectrum is then calculated as $\tau_\nu = - \mathrm{ln} (F_\mathrm{obs} / F_\mathrm{cont}$). The optical depth spectra of low and high luminous sources are presented in Fig. \ref{fig5}. The dip beyond 2270\,cm$^{-1}$ is due to emission of the H$_{2}$ S(10) transition.

Our comparison shows that CO$_2$ ice deposited on both bare and ice-covered silicates at temperatures around 70 K (on laboratory timescales) can qualitatively match not only the double-peaked infrared spectral profile of $^{13}$CO$_2$ observed with JWST on high luminous sources as seen in Figure \ref{fig5} (b) but also the shoulders on the low luminous sources in Figure \ref{fig5} (a), regardless of whether the underlying ice layers are apolar (CO, CH$_4$) or polar (pASW, CW, CH$_3$OH and CH$_3$OH:H$_2$O).} The advantage that our experiments have is that each individual component exhibits a distinct double-peaked profile, so the effect can be attributed solely to the grain surface and there is no need to invoke only CO$_2$ mixed in polar ices at high temperatures. Despite the high luminosity of the sources, it is unlikely that the entire environment will reach uniformly elevated temperatures. Furthermore, we also calculated the full width half maximum (FWHM) of HOPS 370 and IRAS 20126 to be $\sim$12.5 and $\sim$13 cm$^{-1}$, respectively. In comparison, the $^{13}$CO$_2$ profile on both silicate and pASW on silicate at 70 K shown in \autoref{fig:13CO2} has a FWHM of $\sim$10.5 cm$^{-1}$. The broadness of the observational spectra therefore indicate that multiple laboratory components are needed for a better spectral match. For the other young embedded protostars observed by \citet{brunken2024} that were not as luminous as HOPS 370 and IRAS 20126, our laboratory profiles of CO$_2$ ices in polar and apolar layers deposited on silicates at temperatures at or below 70 K can be used to nicely fit the profile as well as the shoulders of the observed peaks.


\section{Conclusion}
We observed a significant effect of the underlying amorphous silicate dust grain analog on the CO$_2$ RAIRS profile. When a silicate film was applied to a gold-coated copper disk, it suppressed the LO mode, which typically dominates on gold due to the MSSR, and enhanced the TO mode, which is typically seen in transmission spectra. To isolate the impact of the silicate grain, we also conducted experiments with bare gold surfaces covered by pASW with varying ice thicknesses representing typical ISM coverages. The CO$_2$ spectra on pASW closely matched those on the bare gold surface. Our results suggest that the relaxation of the MSSR occurs only when the metal surface is covered by a thin film of dust grain analog. This approach maintains the high sensitivity of RAIRS while providing spectra more comparable to observational data, an effect not achievable with ice layers alone, as far as water-experiments are concerned.

In the layered ice experiments, where CO$_2$ was deposited on CO, we found that structural orientation toward crystalline ice occurred as early as the deposition temperature of 15 K. When CO$_2$ was deposited on CH$_4$, structural changes in CO$_2$ occurred between 25 and 30 K. During subsequent warming experiments, we observed that even when CO$_2$ was deposited on pre-existing crystalline water ice layers, the structural orientation occurred at higher temperatures compared to amorphous polar and apolar ice layers. Furthermore, compared to an gold surface, silicate dust grains were found to extend the CO$_2$ desorption temperature, which would result in CO$_2$ remaining in the solid state for a longer period on the astrophysical timescales, allowing more time for chemical reactions to occur to form more complex organic molecules. 

In the warming experiments, we also examined the effect of heating on different CO$_2$ RAIRS modes. Notably, for the $^{13}$CO$_2$ stretching mode, we observed a splitting of the feature around 70 K which did not occur on gold. This split feature closely matched recent JWST observations of young and embedded protostars, which show a similarly split isotopologue feature. In the past, the profile was attributed to contributions from CO$_2$ in different ice mixtures at different temperatures. Based on our study, we propose that in dust grain environments, the $^{13}$CO$_2$ feature splits naturally at a certain temperature, independent of the surrounding ice layer, offering an alternative explanation for the observed double peak structure.

\begin{acknowledgments}
This work is based in part on observations made with the NASA/ESA/CSA James Webb Space Telescope. The data were obtained from the Mikulski Archive for Space Telescopes at the Space Telescope Science Institute, which is operated by the Association of Universities for Research in Astronomy, Inc., under NASA contract NAS 5-03127 for JWST. These observations are associated with program \#1802.
This work has been supported by the European Research Council under the Horizon 2020 Framework Program via the ERC Advanced Grant Origins 83 24 28. We would also like to acknowledge funding through the NanoSpace COST action (CA21126 - European Cooperation in Science and Technology) and Vector Stiftung (Project ID P2023-0152). The laboratory data generated for this study, along with select datasets presented in this paper, are available at: https://doi.org/10.5281/zenodo.15728993.
\end{acknowledgments}

\appendix
During warming, the IR peak profile of the $\nu_3$ band of $^{12}$CO$_2$ exhibited additional changes, summarized in Table~\ref{tab2}. The most notable features include shifts in the LO mode and the appearance of shoulder peaks. These shoulders emerge in the spectral region between the LO and TO modes, with their positions and onset temperatures varying depending on the underlying surface.

\begin{table}[htbp]
\renewcommand{\arraystretch}{1.3}
\centering
\small
\begin{tabularx}{\textwidth}{l 
                                >{\raggedright\arraybackslash}X 
                                >{\raggedright\arraybackslash}X 
                                >{\raggedright\arraybackslash}X 
                                >{\raggedright\arraybackslash}X 
                                >{\raggedright\arraybackslash}X 
                                >{\raggedright\arraybackslash}X 
                                >{\raggedright\arraybackslash}X 
                                >{\raggedright\arraybackslash}X 
                                >{\raggedright\arraybackslash}X}
\toprule
\textbf{T (K)} & \textbf{Au} & \textbf{Si} & \textbf{CO} & \textbf{CH$_4$} & \textbf{pASW} & \textbf{CW} & \textbf{CH$_3$OH} & \textbf{CH$_3$OH:H$_2$O} \\
\midrule
30 & & & &shoulder at 2376 & & & & \\
\midrule
35 & &shoulder at 2371 & & & & &shoulders at 2376 and 2371 & \\
\midrule
40 & shoulders at 2375 and 2368 & & & & & & & \\
\midrule
50 & & & &shoulder start to disappear & & & & shoulder at 2376 \\
\midrule
55 & &shoulders shift to 2365 & & & &shoulder at 2364 & & \\
\midrule
60 & LO redshifts & &LO redshifts, shoulder at 2365 & LO redshifts, new shoulders at 2380 and 2371 & & & &  \\
\midrule
65 & shoulders shift to 2380 and 2366 & & & &shoulder at 2376&shoulder disappear  &  & \\
70 &  & LO blueshifts& & & & & & \\
\midrule
75 & & & & &shoulder disappear & & & \\
\midrule
80 & & & & & &  &shoulders disappear & shoulder disappear \\
\bottomrule
\end{tabularx}
\caption{Summary of IR spectral changes observed for the $\nu_3$ band of $^{12}$CO$_2$ across different layered systems and temperatures.}
\label{tab2}
\end{table}
\par
\noindent 

\newpage

\bibliography{sample631}{}
\bibliographystyle{aasjournal}

\end{document}